\documentclass[mathleft]{an}
\usepackage{graphicx}
\usepackage{times}

\newcommand{\EQ}{\begin{equation}}
\newcommand{\EN}{\end{equation}}
\newcommand{\EQA}{\begin{eqnarray}}
\newcommand{\ENA}{\end{eqnarray}}
\newcommand{\Eq}[1]{Eq.~(\ref{#1})}

\renewcommand{\vec}[1]{{{\mbox{\boldmath $#1$}}}}
\newcommand\uvec[1]{{\widehat{\vec{#1}}}}        
\newcommand{\BB}{{\vec{B}}}

\newcommand{\EE}{{\vec{E}}}
\newcommand{\ee}{{\vec{e}}}

\newcommand{\kk}{{\vec{k}}}

\newcommand{\pp}{{\vec{p}}}

\newcommand{\rr}{{\vec{r}}}

\newcommand{\vv}{{\vec{v}}}

\newcommand{\xx}{{\vec{x}}}

\newcommand{\yy}{{\vec{y}}}

\newcommand{\nab}{{\vec{\nabla}}}
\index{gyration!frequency|see{Larmor frequency}}
\index{gyration!radius|see{Larmor radius}}

\begin{document}

\Pagespan{1}{}
\Yearpublication{}%
\Yearsubmission{}%
\Month{}%
\Volume{}%
\Issue{}%

\title{Magnetic fields in the early universe}
\author{Kandaswamy Subramanian \thanks{Corresponding author:
  \email{kandu@iucaa.ernet.in}\newline}
}
\titlerunning{Magnetic fields in the early universe}
\authorrunning{K. Subramanian}
\institute{
IUCAA, Post Bag 4, Pune University Campus, Ganeshkhind, Pune 411 007, India
}

\received{}
\accepted{}
\publonline{}

\keywords{cosmology: theory -- magnetic fields}

\abstract{%
We give a pedagogical introduction to two aspects of 
magnetic fields in the early universe. 
We first focus on how to formulate electrodynamics
in curved space time, defining appropriate
magnetic and electric fields and writing
Maxwell equations in terms of these fields. 
We then specialize to the case of magnetohydrodynamics in the
expanding universe. We emphasize the usefulness of tetrads in this 
context. We then review the generation
of magnetic fields during the inflationary era, deriving
in detail the predicted magnetic and electric spectra
for some models. We discuss 
potential problems arising from back reaction effects
and from the large variation of the coupling constants
required for such field generation.
  }

\maketitle

\section{Introduction}

The universe is magnetized, from scales of planets to the
largest collapsed objects like galaxy clusters. The origin and
evolution of these magnetic fields is a subject
of intense study. Much of the work on magnetic field origin
centers on the idea that small seed magnetic fields are generated 
by purely astrophysical batteries and are subsequently amplified to
the observed levels by a dynamo (cf. Brandenburg and Subramanian, 2005 
for a review). An interesting alternative is that the observed large-scale
magnetic fields are partially a relic field from the
early Universe, which have been further amplified by motions.

We provide here a pedagogical review of two aspects of
magnetohydrodynamics (MHD) in the early universe. In the first
part of the review we focus on how to formulate electrodynamics
in curved space time, especially how to define magnetic and
electric fields and write Maxwell equations in terms of
these fields. This issue may perhaps be well known to relativists
but may not be so familiar to practitioners of MHD.
The perspective provided is that of the author.

In the second part, we review, again in a pedagogical manner, generation
of magnetic fields during the inflationary era.
It is well known that scalar (density or potential)
perturbations and gravitational waves (or tensor perturbations)
can be generated during inflation. Could
magnetic field perturbations also be generated?
Indeed, a large number of mechanisms whereby magnetic fields are generated
in the early universe have been discussed in the literature
(Turner \& Widrow 1988; Ratra 1992; 
Widrow 2002; Giovannini 2007).
These generically involve the breaking of
the conformal invariance of the electromagnetic action, and the
predicted amplitudes are rather model dependent.
Nevertheless, if a primordial magnetic field
is generated, with a present-day strength of $B \sim 10^{-9}$~G
and coherent on Mpc scales,
it can strongly influence early galaxy formation 
(cf. Sethi and Subramanian, 2005), or induce
signals on the cosmic microwave backround radiation (CMBR) 
(cf. Subramanian, 2006 for a review) including CMBR 
non-Gaussianity (cf. Seshadri and Subramanian, 2009; Caprini et al., 2009). 
An even weaker field,
sheared and amplified due to flux freezing,
during galaxy and cluster formation may kick-start 
the dynamo. It is then worth considering if one can generate such
primordial fields.
Our discussion of inflationary generation of magnetic fields
follows the standard literature, but again the
perspective offered is that of the author.
We hope that our pedagogical discussion of these aspects
of MHD in the early universe will be useful to those
entering the subject.

\section{Electrodynamics in curved spacetime}

We discuss to begin with Maxwell equations in a general curved spacetime
and then focus on FRW models. Electrodynamics in curved spacetime
is most conveniently formulated by giving the action for electromagnetic
fields and their interaction with charged particles:
\EQ
S = -\int \sqrt{-g} \ d^4x \left[ 
\frac{1}{16\pi} F_{\mu \nu}F^{\mu \nu}
- 
A_\mu J^\mu \right]
\label{emaction}
\EN
Here $F_{\mu \nu} = A_{\nu;\mu} - A_{\mu;\nu}
= A_{\nu,\mu} - A_{\mu,\nu}$ is the electromagnetic (EM) field tensor, with
$A_\mu$ being the standard electromagnetic 4-potential and $J^\mu$ the
4-current density.
Further, here and below, we use Greek indices $\mu,\nu$ etc 
for spacetime co-ordinates and Roman indices $i, j, k...$
for purely spatial co-ordinates; repeated indices are summed over 
all the co-ordinates. We also adopt units where the 
speed of light $c=1$ and a metric signature $(-, +, +, +)$.
Demanding that the action is stationary under the variation
of $A_\mu$, gives one half of the Maxwell equations,
\EQ
F^{\mu \nu}_{\quad;\nu} = 4\pi J^\mu
\label{max1}
\EN
And from the definition of the electromagnetic field tensor we also
get the other half of the Maxwell equations
\EQ
F_{[\mu\nu  ; \ \gamma]} = F_{[\mu\nu  , \ \gamma]} = 0; \quad
{\rm or} \
^*F^{\mu\nu}_{\quad ;\nu} = 0 \; .
\label{max2}
\EN
The square brackets $[\mu\nu,\gamma]$
means adding terms with cyclic permutations of $\mu,\nu,\gamma$.
In the latter half of \Eq{max2}, we have defined the dual
electromagnetic field tensor
\[
^*F^{\mu \nu} = \frac12 \epsilon^{\mu\nu\alpha\beta}F_{\alpha\beta} \; .
\]
Here $\epsilon^{\mu\nu\alpha\beta}$ is the totally antisymmetric
Levi-Civita tensor,
\[
\epsilon^{\mu \nu \rho \lambda }=\frac{1}{\sqrt{-g}}
{\cal A}^{\mu \nu \rho \lambda }, \quad
\epsilon_{\mu \nu \rho \lambda }=\sqrt{-g}
{\cal A}_{\mu \nu \rho \lambda },
\]
and $g$ is the determinant of the metric tensor.
Further ${\cal A}^{\mu \nu \rho \lambda }$ is
the totally antisymmetric symbol such that ${\cal A}^{0123}=1$ and $\pm 1$ for
any even or odd permutations of $(0,1,2,3)$ respectively.
Note that ${\cal A}_{0123}=-1$.

We would like to cast these equations in terms of
electric and magnetic fields 
(Ellis, 1973, Tsagas, 2005, Barrow, Maartens \& Tsagas, 2007).
In flat spacetime
the electric and magnetic fields are written in terms of
different components of the EM tensor $F_{\mu\nu}$.
This tensor is antisymmetric, thus its diagonal components
are zero and it has 6 independent components, which can be thought
of the 3 components of the electric field and the 3 components of the
magnetic field. The electric field $E^i$ is given by time-space components
of the EM tensor, while the magnetic field $B^i$
is given by the space-space components
\begin{equation}
F^{0i}=E^{i}\quad F^{12}=B^{3}\quad F^{23}=B^{1}\quad F^{31}=B^{2}\;.
\label{elbdef}
\end{equation}
In a general spacetime, to define corresponding electric and magnetic fields
from the EM tensor, one needs to isolate a time direction. This
can be done by using a family of observers who measure the EM fields
and whose four-velocity is described by the 4-vector
\[
u^{\mu } = \frac{dx^\mu}{ds}; \quad u^\mu u_\mu =-1 \; .
\]
Given this 4-velocity field, one can also define the 'projection tensor'
\[
h_{\mu\nu} = g_{\mu\nu} + u_\mu u_\nu \; .
\]
This projects all quantities into the 3-space orthogonal to $u^\mu$
and is also the effective spatial metric for these observers, i.e
\EQ
ds^2 = g_{\mu\nu} dx^\mu dx^\nu = - (u_\mu dx^\mu)^2
+ h_{\mu\nu} dx^\mu dx^\nu \; .
\EN
Using the four-velocity of these observers, the EM fields can be expressed
in a more compact form as a four-vector electric
field $E_\mu $ and magnetic field $B_\mu$ as
\begin{equation}
E_\mu =F_{\mu \nu }u^{\nu} , \quad
B_\mu ={\frac 12}\epsilon _{\mu \nu \rho \lambda }u^{\nu}F^{\rho \lambda }
= {^*F}_{\mu\nu} u^\nu  \ .
\label{ebnudef0}
\end{equation}
From the definition of $E_\mu$ and $B_\mu$, we have
\[
E_\mu u^\mu = 0, \quad B_\mu u^\mu =0 \; ,
\]
Thus the four-vectors $B_\mu$ and $E_\mu$ have purely spatial components
and are effectively 3-vectors in the space orthogonal to $u^\mu$.
One can also invert \Eq{ebnudef0} to write the EM tensor and its dual in
terms of the electric and magnetic fields
\EQ
F_{\mu\nu} =  u_\mu E_\nu - u_\nu E_\mu
+ \epsilon_{\mu\nu\alpha\beta}B^\alpha u^\beta
\label{feb}
\EN
\EQA
^*F^{\alpha\beta} &=& \frac12 \epsilon^{\alpha\beta\mu\nu}F_{\mu\nu}
\nonumber \\
&=&
\epsilon^{\alpha\beta\mu\nu}
u_\mu E_\nu
+ (u^\alpha B^\beta - B^\alpha u^\beta ) \ .
\label{febdual}
\ENA
We can now use the time-like vector $u^\mu$ and the spatial
metric $h^\mu_\nu$ to decompose the Maxwell equations
into timelike and spacelike parts.
Consider projection of \Eq{max2} on $u_\alpha$. We have
\EQ
u_\alpha(^*F^{\alpha\beta}_{\quad ;\beta})= 0
= (u_\alpha \ ^*F^{\alpha\beta})_{;\beta}
- u_{\alpha;\beta} \ ^*F^{\alpha\beta} 
\label{max2u}
\EN
Substituting \Eq{febdual} into \Eq{max2u} we get
\EQA
&&\left[u_\alpha\epsilon^{\alpha\beta\mu\nu} u_\mu E_\nu
+ u_\alpha u^\alpha B^\beta - u^\beta u_\alpha B^\alpha\right]_{;\beta}
\nonumber \\
&&
-u_{\alpha;\beta}\left[\epsilon^{\alpha\beta\mu\nu}u_\mu E_\nu
+ u^\alpha B^\beta - u^\beta B^\alpha \right] = 0
\label{subs1}
\ENA
The 1st, 3rd and 5th terms are zero because respectively
$u_\alpha\epsilon^{\alpha\beta\mu\nu} u_\mu =0$, $u_\alpha B^\alpha=0$
and $u_{\alpha;\beta}u^\alpha =0$.
The rest of the terms give
\EQ
B^\beta_{;\beta} - B^\beta \dot{u}_\beta
+ u_{\alpha;\beta}\epsilon^{\alpha\beta\mu\nu}u_\mu E_\nu 
= 0 \; ,
\label{divb0}
\EN
where we have defined the acceleration 4-vector,
\[
\dot{u}_\beta = u^\alpha u_{\beta;\alpha}
\]
as the directional derivative of $u^\beta$, along the timelike direction
specified by $u^\alpha$. The covariant velocity gradient tensor,
$u_{\alpha;\beta}$,
can be decomposed into, shear, expansion, vorticity and
acceleration parts in the following manner:
We first write
\EQA
u_{\alpha;\beta} &=& \delta^\mu_\alpha u_{\mu;\nu}
[\delta^\nu_\beta + u^\nu u_\beta]
-\delta^\mu_\alpha u_{\mu;\nu} u^\nu u_\beta
\nonumber \\
&=& h^\mu_\alpha h^\nu_\beta u_{\mu;\nu}
- \dot{u}_\alpha u_\beta
\nonumber \\
&=& \Theta_{\alpha\beta}
+\omega_{\alpha\beta}
- \dot{u}_\alpha u_\beta
\label{sxva}
\ENA
In the last line, we have decomposed the spatial
part of the co-variant derivative $h^\mu_\alpha h^\nu_\beta u_{\mu;\nu}$,
into symmetric expansion tensor $\Theta_{\alpha\beta}$ and an antisymmetric
vorticity tensor $\omega_{\alpha\beta}$. Note that both the expansion
and vorticity tensors are purely spatial, in the sense that
$\Theta_{\alpha\beta} u^\beta =0 = \omega_{\alpha\beta} u^\beta$.
One can further split $\Theta_{\alpha\beta}$ into its trace and trace free part.
\[
\Theta_{\alpha\beta} =
\sigma_{\alpha\beta} +\frac13 \Theta h_{\alpha\beta}
\]
where $\Theta = \Theta^\alpha_\alpha = u^\alpha_{;\alpha}$ is called
the expansion scalar, and $\sigma_{\alpha\beta}$ is called
the shear tensor. Note that the shear tensor satisfies,
$\sigma^\alpha_\alpha = 0$, and $\sigma_{\alpha\beta} u^\beta =0$,
that is it is traceless and also purely spatial.
Thus we have
\EQ
u_{\alpha;\beta} = \sigma_{\alpha\beta} +\frac13 \Theta h_{\alpha\beta}
+ \omega_{\alpha\beta}- \dot{u}_\alpha u_\beta \; .
\label{usplit}
\EN
Only the antisymmetric part of $u_{\alpha;\beta}$ contributes when we substitute
\Eq{usplit} into \Eq{divb0}. Further simplification can be made
by defining the vorticity vector,
\[
\omega^\nu = -\frac12 \omega_{\alpha\beta}\epsilon^{\alpha\beta\mu\nu}u_\mu
\]
and the spatial projection of the covariant derivative
\[
D_\beta B^\alpha = h^\mu_\beta h^\alpha_\nu B^\nu_{;\mu} \; .
\]
Then we have
\EQA
D_\beta B^\beta &=& h^\mu_\beta h^\beta_\nu B^\nu_{;\mu}
= h^\mu_\nu B^\nu_{;\mu}
= (\delta^\mu_\nu + u^\mu u_\nu) B^\nu_{;\mu}
\nonumber \\
&=& B^\mu_{;\mu} - u^\mu u_{\nu;\mu} B^\nu \; .
\ENA
Thus \Eq{divb0} becomes
\EQ
D_\beta B^\beta = 2 \omega^\beta E_\beta \; .
\label{divbeq}
\EN
This equation generalizes the flat space equation
$\nab\cdot\BB =0$, to a general curved spacetime. We see
that $2 \omega^\beta E_\beta$ acts as an effective magnetic
charge, driven by the vorticity of the relative motion
of the observers measuring the electromagnetic field.

Now turn to the spatial projection of
of \Eq{max2} on $h^\kappa_\alpha$. We have
\EQA
h^\kappa_\alpha(^*F^{\alpha\beta}_{\quad ;\beta}) &=& 
h^\kappa_\alpha\left[\epsilon^{\alpha\beta\mu\nu} u_\mu E_\nu
+ u^\alpha B^\beta - u^\beta B^\alpha\right]_{;\beta}
\nonumber \\
&=& 0
\label{max2h}
\ENA
where we have again substituted \Eq{febdual} for the dual EM tensor.
Expanding out the covariant derivative in \Eq{max2h} we get
\EQA
&&h^\kappa_\alpha\left[\epsilon^{\alpha\beta\mu\nu} (u_{\mu;\beta} E_\nu
+ u_\mu E_{\nu;\beta})
+ u^\alpha_{;\beta} B^\beta + u^\alpha B^\beta_{;\beta} \right]
\nonumber \\
&-&h^\kappa_\alpha\left[u^\beta_{;\beta} B^\alpha + \dot{B}^\alpha \right] = 0
\label{subs2}
\ENA
Using $h^\kappa_\alpha = (\delta^\kappa_\alpha + u^\kappa u_\alpha)$
above, only the contribution of $\delta^\kappa_\alpha$ survives
in the 2nd term (since $\epsilon^{\alpha\beta\mu\nu} u_\alpha u_\mu =0$),
3rd term (since $u_\alpha u^\alpha_{;\beta} = 0$) and the 5th term (since
$u_\alpha B^\alpha =0$),
while the 4th vanishes because $h^\kappa_\alpha u^\alpha=0$.
%
Thus we have from the remaining terms,
\EQA
h^\kappa_\alpha \epsilon^{\alpha\beta\mu\nu} u_{\mu;\beta} E_\nu
&+&  \epsilon^{\kappa\beta\mu\nu}u_\mu E_{\nu;\beta}
\nonumber \\
&+& u^\kappa_{;\beta} B^\beta - u^\beta_{;\beta} B^\kappa
- h^\kappa_\alpha \dot{B}^\alpha = 0
\label{inter3}
\ENA
The first term can be simplified using \Eq{usplit}.
Due to the antisymmetry of the Levi-Cevita tensor,
the symmetric parts in \Eq{usplit} drop out and we have,
\EQA
h^\kappa_\alpha \epsilon^{\alpha\beta\mu\nu} u_{\mu;\beta} E_\nu
&=& h^\kappa_\alpha \epsilon^{\alpha\beta\mu\nu} E_\nu \omega_{\mu\beta}
\nonumber \\
&-& h^\kappa_\alpha \epsilon^{\alpha\beta\mu\nu} \dot{u}_{\mu}u_\beta E_\nu \; .
\label{van}
\ENA
The first term on the RHS of the above equation vanishes, i.e
$ h^\kappa_\alpha \epsilon^{\alpha\beta\mu\nu} E_\nu \omega_{\mu\beta}
=0$. To see this it is convenient to define 
a 3-d fully antisymmetric tensor 
\[
\bar{\epsilon}^{ \ \kappa\beta\nu}
= \epsilon^{\kappa\beta\nu\mu}u_\mu \; .
\]
Note that this tensor is also purely spatial in the sense that
$\bar{\epsilon}^{ \ \kappa\beta\nu} u_\nu
= \bar{\epsilon}^{ \ \kappa\beta\nu} u_\beta
= \bar{\epsilon}^{ \ \kappa\beta\nu} u_\kappa = 0$,
a result which follows from the anti-symmetry of
$\epsilon^{\kappa\beta\nu\mu}$.
We can then write the 4-d Levi-Civita tensor as
\EQ
\epsilon^{\kappa\beta\nu\mu}
= 2\left( u^{[\kappa}\bar{\epsilon}^{ \ \beta] \nu\mu}
- \bar{\epsilon}^{ \ \kappa\beta [\nu} u^{\mu]} \right) \; , 
\label{4to3}
\EN
where we use the notation, $A^{[\alpha\beta]}
=(A^{\alpha\beta} - A^{\beta \alpha})/2$.
The first term in \Eq{van} then becomes,
\EQA
&&h^\kappa_\alpha \epsilon^{\alpha\beta\mu\nu} E_\nu \omega_{\mu\beta}
= 
h^\kappa_\alpha E_\nu \omega_{\mu\beta} \times 
\nonumber \\
&&
[u^{\alpha}\bar{\epsilon}^{ \beta \nu\mu}
- u^{\beta}\bar{\epsilon}^{ \alpha \nu\mu}
- \bar{\epsilon}^{ \alpha\beta \nu} u^{\mu}
+ \bar{\epsilon}^{ \alpha\beta \mu} u^{\nu}]
=0 \; .
\label{inden}
\ENA 
The last equality follows from the fact that
$h^\kappa_\alpha u^\alpha =0$, $E_\nu u^\nu=0$ and
$\omega_{\mu\beta} u^\beta = \omega_{\mu\beta} u^\mu =0$.
\footnote{
A more physical way of  
seeing that the first term in
\Eq{van} vanishes is to note that
$E_\nu$ and $\omega_{\mu\beta}$ are purely spatial
tensors, orthogonal to the `time' direction specified by $u^\alpha$.
Thus only the time like component of 
$P^\alpha = \epsilon^{\alpha\beta\mu\nu} E_\nu \omega_{\mu\beta}
$ is non-zero. This implies however that $ h^\kappa_\alpha P^\alpha =0$, thus
proving the required result.
}
Therefore we have
\EQA
h^\kappa_\alpha \epsilon^{\alpha\beta\mu\nu} u_{\mu;\beta} E_\nu
&=& - h^\kappa_\alpha \epsilon^{\alpha\beta\mu\nu} \dot{u}_{\mu}u_\beta E_\nu
\nonumber \\
&=& - \epsilon^{\kappa\beta\mu\nu} \dot{u}_{\mu}u_\beta E_\nu \; .
\label{vanish}
\ENA
The second term of \Eq{inter3} can be written more transparently
by defining 
a `Curl' operator
\[
{\rm Curl}(E^\kappa) = \bar{\epsilon}^{\kappa\beta\nu} E_{\nu;\beta}
\]
Thus $\epsilon^{\kappa\beta\mu\nu}u_\mu E_{\nu;\beta} = - {\rm Curl}(E^\kappa)$.
The third and fourth terms in \Eq{inter3} can also combined and rewritten using \Eq{usplit}
to give,
\EQA
u^\kappa_{;\beta} B^\beta - u^\beta_{;\beta} B^\kappa
&=& \left[u^\kappa_{;\beta} - \Theta \delta^\kappa_\beta \right]B^\beta
\nonumber \\
&=& \left[\sigma^\kappa_\beta +\omega^\kappa_\beta - \frac23 \Theta \delta^\kappa_\beta \right]B^\beta
\ENA
Note that the acceleration term in \Eq{usplit} does not contribute above
because $u_\beta B^\beta = 0$.
Putting all the above results together,
\Eq{inter3} gives the generalization of Faraday law
to curved spacetime,
\EQA
h^\kappa_\alpha\dot{B}^\alpha
&=& \left[\sigma^\kappa_\beta +\omega^\kappa_\beta - \frac23 \Theta \delta^\kappa_\beta \right]B^\beta
\nonumber \\
&-& \bar{\epsilon}^{\kappa\mu\nu} \dot{u}_{\mu} E_\nu
-{\rm Curl}(E^\kappa) \; .
\label{faradayg}
\ENA
The other two Maxwell equations, involving source terms,
can be derived in very similar manner.
Note that if we map $\EE \to -\BB$, and $\BB \to \EE$, then
the dual EM tensor is mapped to the EM tensor, that is
$^*F^{\mu\nu} \to F^{\mu\nu}$. We can use this symmetry
between the EM tensor and its
dual to read off the Maxwell equations involving the
source terms from the source free
equations, \Eq{divbeq} and \Eq{faradayg}. We also note that
in deriving \Eq{divbeq} and \Eq{faradayg}, we changed the sign of
all the terms appearing in \Eq{subs1} and \Eq{subs2}. Thus
mapping $\EE \to -\BB$, and $\BB \to \EE$ in \Eq{divbeq} and \Eq{faradayg}
respectively, and also changing the sign of the source term
$4\pi J^\mu \to -4\pi J^\mu$, the Maxwell equations
$F^{\mu\nu}_{\quad ;\nu} = 4\pi J^\mu$,
in terms of the $E^\mu$ and $B^\mu$ fields, become
\EQ
D_\beta E^\beta = 4\pi \rho_q - 2 \omega^\beta B_\beta  \; ,
\label{diveeq}
\EN
\EQA
h^\kappa_\alpha\dot{E}^\alpha
&=& \left[\sigma^\kappa_\beta +\omega^\kappa_\beta - \frac23 \Theta \delta^\kappa_\beta \right]E^\beta
\nonumber \\
&+& \bar{\epsilon}^{\kappa\mu\nu} \dot{u}_{\mu} B_\nu
+ {\rm Curl}(B^\kappa) - 4\pi j^\kappa \; .
\label{amphereg}
\ENA
Here we have defined the charge and 3-current densities as perceived by the
observer with 4-velocity $u^\alpha$ by projecting the 4-current density
$J^\mu$, along $u^\alpha$ and orthogonal to $u^\alpha$. That is
\[
\rho_q = -J^\mu u_\mu  \; , \quad j^\kappa = J^\mu h^\kappa_\mu.
\]
Note that $j^\kappa u_\kappa = 0$.
To do MHD in the expanding universe,
we also need the relativistic generalization to Ohm's law. This is given by
\EQA
h^\alpha_{(f) \beta} J^\beta &=& \sigma F^{\alpha\beta} w_\beta, \quad
{\rm or} \nonumber \\
J^\alpha &=& \rho_{(f) q} w^\alpha
+ \sigma E_{(f)}^\alpha \; .
\label{relohm}
\ENA
Here the symbol $(f)$ stands for a fluid variable, that is
$w^\alpha$ is the mean 4-velocity of the fluid,
$h^\alpha_{(f) \beta}= (\delta^\alpha_\beta + w^\alpha w_{\beta})$
and $E_{(f)}^\alpha = F^{\alpha\beta} w_{\beta}$ is the electric
field as measured in the fluid rest frame. Also
$\rho_{(f) q}$ and $\sigma$ are the fluid charge densities and
conductivity as measured in its rest frame.
Note that the fluid 4-velocity $w^\alpha$,
need not be the 4-velocity $u^\alpha$
of the family of fundamental observers used to define the EM fields
in Maxwell equations; indeed the conducting fluid will in general
have a peculiar velocity in the rest frame of the fundamental observers.

Further discussion on electrodynamics in curved spacetime (using the 3+1
formalism) and how the different parts of the spacetime geometry affect
the EM field can be found in Tsagas (2005). We now specialize to
case of the expanding universe. 

\subsection{Electrodynamics in the expanding universe}

Let us now consider Maxwell equations for the particular case
of the expanding universe, with the metric that of  
spatially flat Friedmann-Robertson-Walker (FRW) spacetime,
\EQ
ds^2 = -dt^2 + a^2(t) \left[ dx^2 + dy^2 + dz^2\right].
\label{frwmet}
\EN
Here $t$ is the proper time as measured by the fundamental observers
of the FRW universe, while $x,y,z$ are co-moving spatial co-ordinates.
The expansion of the universe is determined by the scale factor $a(t)$,
and $H(t) = \dot{a}/a$ is the Hubble expansion rate (we have also defined
$\dot{a} = da/dt$).
We choose $u^\alpha$ corresponding
to the fundamental observers of the FRW spacetime, that is
$u^\alpha \equiv (1,0,0,0)$. For such a choice and in
the FRW spacetime, we have
\EQ
\dot{u}^\alpha = 0, \quad \omega_{\alpha\beta}=0, \quad
 \sigma_{\alpha\beta} =0, \quad
\Theta = 3 \frac{\dot{a}}{a}.
\label{uchar}
\EN
Further, we can simplify
$h^\kappa_\alpha\dot{B}^\alpha$ as follows:
\EQA
h^\kappa_\alpha\dot{B}^\alpha &=&
(\delta^\kappa_\alpha + u^\kappa u_\alpha) u^\gamma B^\alpha_{;\gamma}
\nonumber \\
&=& u^\gamma B^\kappa_{;\gamma} 
+ u^\kappa u^\gamma[(u_\alpha B^\alpha)_{;\gamma} - u_{\alpha;\gamma}]
\nonumber \\
&=& u^\gamma B^\kappa_{;\gamma}
\ENA
Thus the Maxwell equations reduce to,
\EQA
&& B^\beta_{;\beta} = 0, \qquad
E^\beta_{;\beta} = 4\pi \rho_q,
\nonumber \\
&& u^\gamma B^\kappa_{;\gamma}
+ \frac23 \Theta B^\kappa =
-{\rm Curl}(E^\kappa),
\nonumber \\
&&
u^\gamma E^\kappa_{;\gamma}
+ \frac23 \Theta E^\kappa =
{\rm Curl}(B^\kappa) - 4\pi j^\kappa \;. 
\label{maxfrw}
\ENA
In the spatially flat FRW metric the connection co-efficients
take the form
\EQA
&&\Gamma^0_{0 0} =0 = \Gamma^0_{0 i} = \Gamma^{i}_{j k},
\quad \Gamma^0_{i j} = \delta_{ij} a \dot{a},
\nonumber \\
&& \Gamma^i_{0 j} = \delta_{i j} \frac{\dot{a}}{a}.
\label{connection}
\ENA
Using these \Eq{maxfrw} can be further simplified as follows:
\EQA
&& \frac{\partial B^i}{\partial x^i} = 0,
\qquad 
\frac{\partial E^i}{\partial x^i} =
4\pi \rho_q,
\nonumber \\
&&
\frac{1}{a^3} \frac{\partial}{\partial t} \left[a^3 B^i\right]
= -\frac{1}{a} \epsilon^*_{ilm} \frac{\partial E^l}{\partial x^m},
\nonumber \\
&&
\frac{1}{a^3} \frac{\partial}{\partial t} \left[a^3 E^i\right]
= \frac{1}{a} \epsilon^*_{ilm} \frac{\partial B^l}{\partial x^m}
- 4 \pi j^i \;.
\label{maxsimp}
\ENA
Here we have defined the 3-d fully antisymmetric symbol
$\epsilon^*_{ijk}$, with $\epsilon^*_{123} = 1$.
These equations resemble the flat spacetime Maxwell equations
except for the presence of the scale factor $a(t)$.

The electric and magnetic field 4-vectors we have used above
are referred to a co-ordinate basis, where the spacetime metric
if of the FRW form. They have the following curious property.
Consider for example the case when the plasma in the universe has
no peculiar velocity, that is $w^\alpha = u^\alpha$,
and also highly conducting 
with $\sigma \to \infty$. Then from \Eq{relohm}, we have 
$E_{(f)}^\alpha = 0 = E^\alpha$, and from Faraday's law
in \Eq{maxsimp}, $B^i \propto 1/a^3$.
There is however a simple result derivable in flat space time that
in a highly conducting fluid, the magnetic flux through
a surface which co-moves with the fluid is constant.
Since in the expanding universe all proper surface areas
increase as $a^2(t)$, one would expect the strength
of a 'proper' magnetic field to go down with expansion as $1/a^2$.
This naively seems to be at variance with
the fact that $B^i \propto 1/a^3$ and $B_i = g_{i\mu} B^\mu \propto 1/a$. 
There are two comments to made at this stage: First, if we define the magnetic field
amplitude, say $B$, by looking at the norm of the four vector $B^\mu$,
that is let $B^2 = B^\mu B_\mu = B^iB_i \propto 1/a^4$, then we do get
$B \propto 1/a^2$. This procedure however does not appear
completely satisfactory as one would prefer to deal with the field
components themselves. Another possibility is to refer 
all tensor quantities to a set of orthonormal basis vectors, referred
to as tetrads.

Any observer can be thought to be carrying along
her/his world line a set of four orthonormal vectors 
${\bf e}_{(a)}$, where $a=0,1,2,3$, which satisfy the relation
\EQ
g_{\mu\nu} e^\mu_{(a)} e^\nu_{(b)} = \eta_{ab},
\quad \eta^{a b} e^\mu_{(a)} e^\nu_{(b)} = g^{\mu \nu}
\label{tetradprop}
\EN
Here $\eta_{a b}$ has the form of the flat space-time metric.
We choose the observer's 4-velocity itself to be the tetrad with $a=0$, i.e
$e^\mu_{(0)} = u^\mu$. The other three tetrads are orthogonal to the
observer's 4-velocity. In the present case, we consider the observer to
be the fundamental observer of the FRW space time, and the
components of the tetrads, which satisfy \Eq{tetradprop} are given by
\[
e^\mu_{(0)} = \delta^\mu_0, \quad e^\mu_{(i)} = \frac{1}{a} \delta^\mu_i, \quad i = 1,2,3
\]
The metric $\eta_{a b}$ can also be used to
raise and lower the index of the tetrad to define
$e^{\mu (a)} = \eta^{a b} e^\mu_{(b)}$.
Note that the fundamental observers move along geodesics, and
as we noted earlier, do not have either relative
acceleration or rotation.
Such observers parallel transport their tetrad along their
trajectory, i.e $u^\mu e^\alpha_{(a);\mu} = 0$, as can be easily
checked dy direct calculation using the connection co-efficients
given in \Eq{connection}.
The magnetic and electric field components can now be
represented as its projection along the four orthonormal tetrads using,
\EQ
\bar{B}^{a} = g_{\mu\nu} B^\mu e^{\nu (a)}, \quad \bar{E}^{a} = g_{\mu\nu} E^\mu e^{\nu (a)}, 
\label{EBtetrad}
\EN
which gives 
\EQA
\bar{B}^0 &=& 0, \quad \bar{E}^0 = 0, \nonumber \\
\bar{B}^a &=& a(t) B^a,  \quad \bar{E}^a = a(t) B^a, 
\ {\rm for} (a=1,2,3).
\label{EBcomp}
\ENA
Note that $\bar{B}^{a}, \bar{E}^{a}$ are co-ordinate scalars, but the set of four scalars
$\bar{B}^{a}$ is still a vector as far as local Lorentz transformation is concerned
(which preserves the orthonormality conditions in \Eq{tetradprop}). If we
define $\bar{B}_{a} = \eta_{ab} \bar{B}^b$, then numerically $\bar{B}^{i} = \bar{B}_i$ and
$\bar{B}^{0} = - \bar{B}_0 = 0$.
Similar relations  obtain for the electric field components.
In the FRW universe, as $B^i \propto 1/a^3$, we see that
$\bar{B}^{i} = \bar{B}_i \propto 1/a^2$, as one naively expects
from flux freezing of the magnetic field. Thus the magnetic field
components projected onto the orthonormal tetrads seem to be
the natural quantities to be used as the 'physical' components of the
magnetic field. 
Note that this is similar to using the Cartesian components of
a vector as the physical components in 3 dimensional vector analysis.

There is an additional feature of using tetrads which is
of particular interest. 
Given the set of tetrads one can set up a local
co-ordinate system around any event ${\cal P}$ by using geodesics
emanating from ${\cal P}$ and whose tangent vectors at ${\cal P}$ are the
four tetrads ${\bf e}_{(a)}$. This co-ordinate frame, is a locally inertial frame; that
is the spacetime is locally flat with the metric in the form of
$\eta_{a b}$ and the connection co-efficients in these co-ordinates
vanishing
(see section 13.6
of Misner, Thorne \& Wheeler, 1971
(MTW) for a proof).
In fact such a co-ordinate system can be set up all along
the world line of the fundamental observer, and are called
then Fermi-Normal co-ordinates 
(Manasse \& Misner, 1963; MTW; 
Cooperstock, Faraoni and Vollich, 1998).
To leading order, the time direction in this inertial frame,
is the proper time co-ordinate $t$ of the FRW metric and the
space co-ordinates become the proper space co-ordinates $r^i = a (t) x^i$.
(There are second order deviations from these relations 
(cf. Cooperstock, Faraoni and Vollich, 1998)).

The Maxwell equations in such a locally inertial frame take a particularly
transparent form. They can be derived from \Eq{maxfrw} by adopting the flat space metric
and replacing the covariant derivative with an ordinary derivative.
As the metric is locally flat, the curl operator also reduces to the
ordinary curl with respect to $r^i$. Importantly $\Theta$ being a scalar
is invariant under co-ordinate transformations, and so still $\Theta = 3\dot{a}/a$.
The field components are the same as those defined above using the tetrads.
This is because the tetrads become the co-ordinate basis vectors 
in the Fermi-Normal co-ordinates.
The Maxwell equations then become,
\EQA
&& \frac{\partial \bar{B}^i}{\partial r^i} = 0,
\qquad 
\frac{\partial (a^2 \bar{E}^i)}{\partial r^i} =
4\pi \rho_q a^2,
\nonumber \\
&&
\frac{\partial}{\partial t} \left[a^2 \bar{B}^i\right]
= - \epsilon^*_{ilm} \frac{\partial (a^2 \bar{E}^l)}{\partial r^m},
\nonumber \\
&&
\frac{\partial}{\partial t} \left[a^2 \bar{E}^i\right]
= \epsilon^*_{ilm} \frac{\partial (a^2 \bar{B}^l)}{\partial r^m}
- 4 \pi \bar{j}^i a^2 \;.
\label{max_inertial}
\ENA

In the absence of charges and currents, \Eq{max_inertial} has electromagnetic wave
solutions, with the amplitude of the electric and magnetic fields
decaying with expansion as $1/a^2$. In the presence of a conducting medium,
one has to again supplement these Maxwell equations with the Ohm's law.
In the limit of non-relativistic fluid velocity $\vv$, this
again reduces to
\EQ
\bar{j}^i = \rho_q v^i + \sigma \left[ \bar{E}^i + \epsilon^{*ilm} v_l \bar{B}_m \right]
\EN
If we neglect the displacement current and charge density terms,
as valid for a highly conducting medium (cf. Brandenburg and Subramanian, 2005),
the induction equation becomes
\EQ
\frac{\partial}{\partial t} \left( \BB a^2 \right)
= \nab_{\rr} \times \left [ \vv \times (\BB a^2) - \eta \nab \times (\BB a^2) \right]
\label{expind}
\EN
where we have defined the vector $\BB \equiv (\bar{B}^1, \bar{B}^2, \bar{B}^3)$.
Thus we see that in the absence of resistivity ($\eta =0$)
or peculiar velocities ($\vv =0$), the magnetic field defined in the
local inertial frame, decays with expansion factor
as $\BB \propto 1/a^2$. As pointed out above, this decays is as expected, when the magnetic
flux is frozen to the plasma, since all proper areas in the FRW spacetime
increase with expansion as $a^2$.
This completes our pedagogical discussion of doing magnetohydrodynamics
in curved spacetime and in particular the expanding universe.
We turn to a pedagogical discussion of primordial magnetic field
generation during the inflationary era.

\section{Magnetic field generation during inflation}

The early universe is supposed to have gone through an
epoch of accelerated expansion referred to as inflation.
Inflation can provide a solution to  
several problems of standard big bang cosmology, 
one of them being to explain the origin of perturbations
which eventually led to all the structures that we see.
We refer the reader to several standard textbooks 
for a discussion of the inflationary paradigm (Liddle and Lyth, 1999;
Mukhanov, 2005; Padmanabhan, 2002) and recent reviews
(Bassett, Tsujikawa, Wands, 2006;
Sriramkumar, 2009).
Inflation provides several ideal conditions for the generation
of primordial fields with large coherence scales (Turner \& Widrow, 1988).
First the rapid expansion in the inflationary era provides the kinematical
means to produce fields correlated on very large scales by just the
exponential stretching of wave modes. Also vacuum fluctuations
of the electromagnetic (or more correctly the hypermagnetic) field
can be excited while a mode is within
the Hubble radius and could be transformed
to classical fluctuations as it transits outside the Hubble radius.
Finally, during inflation any existing charged particle densities
are diluted drastically by the expansion, so that the universe is
not a good conductor; thus magnetic flux conservation then does not
exclude field generation from a zero field.
Most of the models for magnetic field generation during inflation
take the field to be described by the action of an abelian gauge field, and
have not considered the action obtained from for example the Electro-Weak 
or some Grand Unified theory.
For simplicity we shall also adopt this approach below.

There is one major difficulty, which arises when one considers
magnetic field generation during inflation. This is because
the standard electromagnetic
action is conformally invariant, and the universe metric is conformally flat.
Consider again the electromagnetic action
\EQA
S &=& -\int \sqrt{-g} \ d^4x \ \frac{1}{16\pi} F_{\mu \nu}F^{\mu \nu}
\nonumber \\
&=& - \int \sqrt{-g} \ d^4x \ \frac{1}{16\pi}
g^{\mu \alpha}g^{\nu \beta} F_{\mu \nu}F_{\alpha \beta}
\ENA
Suppose we make a conformal transformation of the metric given by
\EQ
g^*_{\mu \nu} = \Omega^2 g_{\mu \nu}
\label{conformal}
\EN
This implies $\sqrt{-g^*} =\Omega^4\sqrt{-g}$ and
$g^{* \mu \alpha}= \Omega^{-2} g^{\mu \alpha}$. Then
taking
\EQ
A_\mu^* = A_\mu \Rightarrow S^* = S.
\label{invariant_action}
\EN
Thus the action of the free electromagnetic field is
invariant under conformal transformations.
Note that the FRW models are conformally flat;
that is the FRW metric can be written as
$g^{FRW}_{\mu \nu} = \Omega^2 \eta_{\mu \nu}$, where
$\eta_{\mu \nu}$ is the Minkowski, flat space metric.
As we show below explicitly for a universe with flat spatial sections,
this implies that one can transform
the electromagnetic wave equation into its flat space version.
It turns out that one cannot then amplify electromagnetic
wave fluctuations in such a FRW universe and
the field then always decreases
\footnote{Interestingly slower decay can occur for modes comparable to the
curvature scale, in open FRW models due to coupling to the
spatial curvature (Barrow and Tsagas, 2008).
}
with expansion as $1/a^2(t)$.

Therefore mechanisms for magnetic field generation require
the breaking of conformal invariance of the electromagnetic action,
which changes the above behaviour to $B \sim 1/a^{\epsilon}$ with typically
$\epsilon \ll 1$ for getting a strong field.
A multitude of ways have been considered for breaking conformal invariance  of the
EM action during inflation. Some of them are illustrated in the
action below:
\EQA
S &=& \int \sqrt{-g} \ d^4x \ [ -f^2(\phi,R) \frac{1}{16\pi} F_{\mu \nu}F^{\mu \nu}
-b R A^2 \nonumber \\
&+& g \theta F_{\mu \nu} \tilde{F}^{\mu \nu}
- D_{\mu}\psi (D^\mu \psi)^* \ ]
\label{cibreak}
\ENA
They include coupling of EM action to scalar fields ($\phi$) like the
inflaton and the dilaton, coupling to curvature invariants ($R$),
coupling to a pseudo-scalar field like the axion ($\theta$),
having charged scalar fields ($\psi$) and so on.
If conformal invariance of the EM action
can indeed be broken, the EM wave can amplified from vacuum fluctuations,
as its wavelength increases from sub-Hubble to super-Hubble scales.
After inflation ends, the universe reheats and leads to the production
of charged particles leading to a dramatic increase in the plasma
conductivity. Then the electric field $\EE$ would get
shorted out while the magnetic field $\BB$ of the EM wave gets frozen in.
This is the qualitative picture for the generation
of primordial fields during the inflationary era.

There is however another potential difficulty;
since $a(t)$ is almost exponentially
increasing during slow roll inflation, the predicted field amplitude,
which behaves say as $B \sim 1/a^{\epsilon}$ is
exponentially sensitive to any changes of the parameters of the model
which affects $\epsilon$. Therefore models of magnetic
field generation can lead to fields as large as $B\sim 10^{-9}$ G
(as redshifted to the present epoch) down to
fields which are much smaller than that required for even seeding 
the galactic dynamo. For example in  model considered by 
Ratra (1992) with $ f^2(\phi) \sim e^{\alpha\phi}$,
with $\phi$ being the inflaton, one gets $B\sim 10^{-9}$ to $10^{-65}$ G,
for $\alpha \sim 20-0$.
Note that the amplitude of scalar perturbations generated
during inflation is also dependent on the parameters of the theory and
has to be fixed by hand. But the sensitivity to parameters seems
to be stronger for magnetic fields than for scalar perturbations due
to the above reason. Nevertheless one may hope that there would
arise a theory where the parameters are naturally such as to produce
interesting primordial magnetic field strengths. We describe
below one framework for magnetic field generation during
inflation, keeping the discussion quite general without
specifying any specific inflation model. A nice treatment
of inflationary generation of magnetic fields, which
we follow to some extent, is given by Martin and Yokoyama (2008),
where some specific models are also discussed.

\subsection{Quantizing the EM field}

Let us assume that the scalar field $\phi$ in \Eq{cibreak}
is the field responsible for inflation and also assume that this
is the sole term which breaks the conformal invariance of the
electromagnetic action. The total action is given by
\EQA
S &=& -\frac{1}{16\pi} \int {\rm d}^4x \sqrt{-g}
\left[ g^{\alpha \beta }g^{\mu \nu} f^2\left(\phi \right)
F_{\mu \alpha }F_{\nu \beta}\right] \nonumber \\
&&- \int d^4x \sqrt{-g} \left[ \frac12 g^{\mu \nu} \partial_\mu\phi
\partial_\nu\phi + V(\phi) \right]
\label{bgen}
\ENA
Maxwell equation now become
$[f^2F^{\mu\nu}]_{;\nu}= 0$, or
\EQ
\frac{1}{\sqrt{-g}}\frac{\partial}{\partial x^\nu}
\left[\sqrt{-g} g^{\alpha \beta } g^{\mu \nu} f^2\left(\phi \right)
F_{\mu \alpha }F_{\nu \beta}\right] 
= 0
\label{max_f}
\EN
The scalar field satisfies
\EQ
\frac{1}{\sqrt{-g}}\frac{\partial}{\partial x^\nu}
\left[ \sqrt{-g} g^{\mu \nu} \partial_\mu\phi\right]
-\frac{dV}{d\phi} = \frac{f}{8\pi} \frac{df}{d\phi} F_{\mu\nu}F^{\mu\nu}
\EN
We assume that the electromagnetic field is a `test' field
which does not perturb either the scalar field evolution or 
the evolution of the background FRW universe. We take the metric
to be spatially flat with
\EQA
ds^2 &=& -dt^2 + a^2 \left[ dx^2 + dy^2 + dz^2\right]
\nonumber \\
&=& a^2(\eta)\left[-d\eta^2 + dx^2 + dy^2 + dz^2\right]
\label{metric}
\ENA
where $\eta = \int (dt/a) $ is the conformal time.
Throughout the discussion in this section, we use conformal time
($\eta$) and co-moving space $(x,y,z)$ as our four co-ordinates
and all tensors when explicitly specified will be in this co-ordinate frame.
It is convenient to adopt the Coulomb gauge: 
\EQ
A_0(\eta,{\bf x}) = 0, \qquad {\partial}_jA^j(\eta,{\bf x}) =0 \; .
\label{Coul_gauge}
\EN
In this case the time component of 
\Eq{max_f} becomes a trivial identity, while
the space components give
\EQ
A_i{''} 
+ 2\frac{f'}{f}A_i' - a^2\partial_j\partial^j A_i = 0
\label{Aevol}
\EN
where we have defined 
$\partial^j = g^{jk}\partial_k = a^{-2}\delta^{jk}\partial_k$,
and a prime denotes derivative with respect to $\eta$.
In fact $a^2\partial_j\partial^j$ is the usual spatial 
$\nabla^2$ operator with respect to the co-moving spatial co-ordinates.

We can also use \Eq{ebnudef0} to write the electric and magnetic
fields in terms of the vector potential. Note that the
four velocity of the fundamental observers used to define 
these fields is now given by $u^\mu \equiv (1/a, 0,0,0)$.
The time components of $E_\mu$ and $B_\mu$ are zero, while 
the spatial components are given by
\EQ
E_i = -\frac{1}{a} A_i' ; 
\quad B_i = \frac{1}{a}{\epsilon}_{ijk}^*\delta^{jl}\delta^{km}
(\partial_l A_m)
\label{EBinA}
\EN
Note that these spatial components are the same as when $t$
is used instead of $\eta$ as the time co-ordinate.
This is because the transformation $t\to \eta$ is independent of
$\xx$, and we have not transformed
the space co-ordinates.
For a constant $f$ \Eq{Aevol} shows that $A_i$ simply satisfies
the usual wave equation in $\eta$ and $\xx$ co-ordinates, whose
solutions are plane waves with constant amplitude.
Then the amplitude of $B_i$ then scales as $1/a$, while that of $B^i$ scales as $1/a^3$
and so the amplitude of $\bar{B}^a$ scales as $1/a^2$ as before.

We would like quantize the electromagnetic field in the FRW background.
For this we treat $A_i$ as the co-ordinate, and find the conjugate
momentum $\Pi^i$, 
by varying the EM Lagrangian density 
${\cal L}_{EM} = -f^2 F_{\mu\nu}F^{\mu\nu}/(16\pi)$,
with respect to $A_i'$. We get
\[
\Pi^i = \frac{\delta{\cal L}_{EM}}{\delta A_i'}
=\frac{1}{4\pi} f^2 a^2 g^{ij} A_j', \quad \Pi_i 
= \frac{1}{4\pi} f^2 a^2 A_i'
\]
To quantize the electromagnetic field, we promote $A^i$ and $\Pi_i$
to operators and impose the canonical commutation relation between them,
\EQA
\left[A^i(\xx,\eta),\Pi_j(\yy,\eta) \right] 
&=& i \int \frac{d^3k}{(2\pi)^3}  \  e^{i\kk\cdot(\xx-\yy)} P^i_j(\kk)
\nonumber \\
&=& i \delta_{\perp \ j}^i(\xx-\yy) \;.
\label{quant1}
\ENA
Here the term $P^i_j(\kk) = (\delta^i_j -\delta_{jm} (k^i k^m/k^2))$
is introduced to ensure that the Coulomb gauge condition is
satisfied and $\delta_{\perp \ j}^i$ is the transverse delta function.
This quantization condition is most simply incorporated
in Fourier space. We expand the vector potential in
terms of creation and annihilation operators, 
$b_\lambda^{\dagger}(\kk)$ and $b_\lambda(\kk)$, with $\kk$ the
co-moving wave vector,
\EQA
&&A^i(\xx,\eta) = \sqrt{4\pi} 
\int \frac{ d^3 k}{(2\pi)^3}\sum _{\lambda =1}^2
{\cal \ee}^i_{\lambda }(\kk) \times
\nonumber \\
&&\left[
b_\lambda(\kk) A(k,\eta)e^{i \kk\cdot\xx }
+ b_\lambda^{\dagger}(\kk) {A}^*(k,\eta)e^{-i\kk\cdot\xx} \right] \;.
\label{creation_exp}
\ENA
Here the index $\lambda=1,2$ and ${\cal \ee}^i_{\lambda }(\kk)$
are the polarization vectors, which form part of  
an orthonormal set of basis four-vectors,
\EQ
{\cal\ee}^\mu_0 = \left( \frac{1}{a}, {\bf 0} \right), 
{\cal\ee}^\mu_\lambda = \left( 0, \frac{\bar{\cal\ee}^i_{\lambda}}{a} \right), 
{\cal\ee}^\mu_{3} = \left( 0, \frac{\uvec{k}}{a}\right) \; . 
\label{orthodef}
\EN
The 3-vectors $\bar{\cal\ee}^i_{\lambda}$ are unit vectors, 
orthogonal to $\kk$ and to each other. The
expansion in terms of the polarization vectors incorporates
the Coulomb gauge condition in Fourier space. It also shows that
the free electromagnetic field has two polarization degrees of freedom.
If we substitute the Fourier expansion given in 
\Eq{creation_exp} into \Eq{Aevol}, we 
find that the Fourier
coefficients $\bar{A} = (aA(k,\eta))$ satisfy,
\EQ
\bar{A}{''} + 2\frac{f'}{f}\bar{A}' + k^2 \bar{A} = 0
\label{barAevol}
\EN
One can also define a new variable 
${\cal A} = a(\eta )f(\eta )A(\eta ,k) $ in order to eliminate
the first derivative term, to get
\EQ
{\cal A}''(\eta ,k)+\left(k^2-\frac{f''}{f}\right){\cal A}(\eta ,k)=0
\label{calAeq}
\EN
We can see that the mode function ${\cal A}$ satisfies the
equation of a harmonic oscillator with a time dependent mass term.
The case $f=1$ corresponds to the standard EM action where ${\cal A}$
oscillates harmonically with time. One needs $f'' >  0$ to have possible
growth of magnetic fields.

The quantization condition given in
\Eq{quant1} is implemented by imposing 
the following commutation relations between the creation and
annihilation operators,
\EQA
\left[ b_\lambda(\kk), b_{\lambda'}^\dagger(\kk')\right] &=&
(2\pi)^3 \ \delta^3(\kk -\kk') \ \delta_{\lambda\lambda'}, \nonumber \\ 
\left[ b_\lambda(\kk), b_{\lambda'}(\kk')\right] &=&
\left[ b_\lambda^\dagger(\kk), b_{\lambda'}^\dagger(\kk')\right] =
0 \;.
\label{bcommutation}
\ENA
We also define the vacuum state $\vert0>$ as one which is annihilated
by $b_\lambda(\kk)$, that is $b_\lambda(\kk)\vert0> = 0$.
Note that the choice of the initial quantum state 
will be decided by the choice of the mode function ${\cal A}$ as below.
To check if \Eq{quant1} is indeed satisfied, we can
substitute the Fourier expansion of $A^i$ and $\Pi_j$ into
the commutator $[A^i,\Pi_j]$, and use the
commutation relations \Eq{bcommutation}. We get
\EQA
\left[A^i(\xx,\eta),\Pi_j(\yy,\eta) \right] 
&=& \int \frac{ d^3 k}{(2\pi)^3}\sum _{\lambda =1}^2
 {\cal \ee}^i_{\lambda }(\kk){\cal \ee}_{j \lambda }(\kk) 
\times 
\nonumber \\
&& e^{i\kk\cdot(\xx-\yy)} W(k,\eta)
f^2 a^2,
\label{intercom}
\ENA
where we have defined the complex Wronskian, 
\[
W(k,\eta) = [ A A'^{*} -A^* A']. 
\]
One can check that the polarization four-vectors
satisfy the completeness relation,
\EQ
\sum _{\lambda =1}^2 {\cal \ee}^i_{\lambda }(\kk){\cal \ee}_{j \lambda }(\kk) 
= P^i_j(\kk)
\label{complete}
\EN
Let us also define the
Wronskian associated with $\bar{A}$, given by
$\bar{W} = [\bar{A}\bar{A}'^{*} -\bar{A}^* \bar{A}']$.
We have $\bar{W} = a^2 W$, and since 
$\bar{A}$ satisfies \Eq{barAevol}, we get
$\bar{W}' = -(2f'/f) \bar{W}$. Integrating this equation
we get $\bar{W} = a^2 W \propto (1/f^2)$.
Substituting the expression for $W$ 
into \Eq{intercom}, and using \Eq{complete}, 
we can verify that
the quantization condition \Eq{quant1} is indeed satisfied,
provided we fix the constant of proportionality in $W$   
such that $W = (i/f^2 a^2)$.

Once we have set up the quantization of the EM field, it
is of interest to ask how the energy density of the EM
field evolves. The energy momentum tensor is given by
varying the EM Lagrangian density with respect to the metric,
\EQA
T_{\mu\nu} &=& -\frac{2}{\sqrt{-g}} \frac{\delta[\sqrt{-g} {\cal L}_{EM}]}
{\delta g^{\mu\nu}}
\nonumber \\
&=& \frac{f^2}{4\pi} \left[ g^{\gamma\beta} F_{\mu\gamma}F_{\nu\beta} 
-g_{\mu\nu} \frac{F_{\alpha\beta}F^{\alpha\beta}}{4}\right]
\label{emtensor}
\ENA
The energy density $T_{\mu\nu}u^\mu u^\nu$ can be written as the sum of a magnetic and
electric contributions. The energy density 
due to the magnetic part of the EM field
is given by
\EQA
T_{\mu\nu}^B u^\mu u^\nu &=& \frac{f^2}{16\pi}[A_{m,i} - A_{i,m}][A_{l,j} - A_{j,l}]
g^{ij} g^{ml} 
\nonumber \\
&=& f^2 \frac{B_iB^i}{8\pi} \;.
\label{magenergy}  
\ENA
Similarly, the energy density due to the electric field is
\EQ
T_{\mu\nu}^E u^\mu u^\nu  = \frac{f^2}{8\pi}[A_i' A_j']
g^{ij} = f^2 \frac{E_iE^i}{8\pi} \; .
\label{elecenergy}  
\EN
We substitute the Fourier expansion of $A_i$ 
into \Eq{magenergy} and \Eq{elecenergy},
and take the expectation value in the vacuum state
$\vert 0 >$. Let us define
\[
\rho_B = <0\vert T_{\mu\nu}^B u^\mu u^\nu \vert0>, \quad 
\rho_E = <0\vert T_{\mu\nu}^E u^\mu u^\nu \vert0> \; .
\] 
 Using the properties
\EQA
b_\lambda(\kk)\vert0> &=& 0, \nonumber \\
<0\vert b_\lambda(\kk) b_{\lambda'}^\dagger(\pp)\vert 0 > &=& 
(2\pi)^3 \delta(\kk - \pp)\delta_{\lambda\lambda'},
\label{vac}
\ENA
we get for the spectral energy densities in the magnetic and electric fields,
\EQ
\frac{d\rho_B}{d\ln{k}} = \frac{1}{2\pi^2} \left(\frac{k}{a} \right)^4
k \left\vert {\cal A}(k,\eta) \right\vert^2 ,
\label{rhoB}
\EN
\EQ
\frac{d\rho_E}{d\ln{k}} = \frac{f^2}{2\pi^2} \frac{k^3}{a^4} 
\left\vert \left[\frac{{\cal A}(k,\eta)}{f} \right]' \right\vert^2 \; .
\label{rhoE}
\EN
Once we have calculated the evolution of the mode function 
${\cal A}(k,\eta)$, the evolution of energy densities in the
magnetic and electric parts of the EM field can be calculated.

\subsection{Evolution of normal modes}

Consider for example a case where the scale factor $a(\eta)$
evolves with conformal time as
\EQ
a(\eta) = a_0 \left\vert \frac{\eta}{\eta_0}\right\vert^{1 + \beta}
\label{scale_evol}
\EN
The case when $\beta = -2$ corresponds to de Sitter space-time of
exponential expansion in cosmic time, or $a(t) \propto \exp{(Ht)}$.
On the other hand for an accelerated power law expansion with 
$a(t) = a_0(t/t_0)^p$ and $p > 1$, integrating $dt = ad\eta$, we have 
\EQA
\eta &=& -\frac{t_0}{a_0(p-1)} \left(\frac{t}{t_0}\right)^{-1/(p-1)},
\nonumber \\
a(\eta) &=& a_0 \left[\frac{-a_0(p-1)\eta}{t_0}\right]^{-p/(p-1)} \; .
\label{expfac}
\ENA
Here we have assumed that $\eta \to 0_{-}$ as $t \to \infty$, such that 
during inflation, the conformal time lies in the range $-\infty < \eta < 0$.
In the limit of $p \gg 1$, one goes over to an almost
exponential expansion with $\beta \to -2 -1/p$.

Let us also consider a model potential where the gauge coupling
function $f$ evolves as a power law,
\EQ
f(\eta) \propto a^\alpha \; .
\label{fevol}
\EN
This could obtain for example for exponential form of $f(\phi)$ and
power law inflation. We then have
\EQ
\frac{f''}{f} = \frac{\gamma(\gamma -1)}{\eta^2},
\quad 
\gamma = \alpha(1+\beta)
\EN
Then the evolution of the mode function ${\cal A}$ is given by
\EQ
{\cal A}''(k,\eta)
+\left(k^2-\frac{\gamma(\gamma -1)}{\eta^2}\right){\cal A}(k,\eta)=0,
\label{Aevol2}
\EN
whose solution can be written interms of Bessel functions,
\EQA
{\cal A} = (-k\eta)^{1/2} &[& \ C_1(k) J_{\gamma -1/2}(-k\eta)
\nonumber \\
&&+ \  C_2(k) J_{-\gamma + 1/2}(-k\eta) \ ],
\label{Asoln}
\ENA
where $C_1(k)$ and $C_2(k)$ are scale-dependent coefficients to be fixed
by the initial conditions.

Let us define the Hubble radius at any epoch as the length scale $R_H= 1/H$
(in units where the speed of light $c=1$).
The initial conditions to determine the constants in
\Eq{Asoln} are specified for each mode (or wavenumber $k$),
when it is deep within the Hubble radius.
That is when the proper scale length associated with the
mode $(k/a)^{-1}$ is much smaller than $R_H$, or $(k/aH) \gg 1$.
For such small scale modes
one assumes that the effects of space-time curvature are negligible and
thus the mode
function goes over to that relevant for the Minkowski space vacuum.
Recall that the expansion rate is given by $H(t) = \dot{a}/a = a'/a^2$.
(Here and below $\dot{A}$ is derivative of $A$ with respect to proper time.)
For the expansion factor given in \Eq{expfac}, we have 
$a'/a = -(p/(p-1))(1/\eta)$, and for $p \gg 1$, $aH \to -1/\eta$.
Thus the ratio the Hubble radius to the proper scale of a perturbation is
given by $(1/H)(a/k)^{-1} = k/(aH) = -k\eta$. A given mode is
therefore within the Hubble radius for $-k\eta > 1$ and outside the Hubble
radius when $-k\eta < 1$.

In the short wavelength limit, $(k/a)/H = (-k\eta) \to \infty$, 
the solutions of \Eq{Aevol2} are simply ${\cal A} \propto \exp{(\pm ik\eta)}$.
The assumption that the gauge field for these modes is in the
Minkowski space vacuum state leads us to pick the 'positive' frequency mode
${\cal A} =c_0 \exp{(- ik\eta)}$, and from the Wronskian
condition that $W = i/f^2a^2$, the constant $c_0$ is fixed to 
$c_0 = 1/\sqrt{2k}$.
Thus we assume
as initial condition that as $(-k\eta) \to \infty$,
\EQ
{\cal A} \to \frac{1}{\sqrt{2k}}  \exp^{-ik\eta}.
\label{initialA}
\EN
This fixes the constants in \Eq{Asoln} to be,
\EQA
C_1(k) &=& \sqrt{\frac{\pi}{4k}} \frac{\exp{(-i\pi\gamma/2)}}{\cos(\pi\gamma)}, 
\nonumber \\
C_2(k) &=& \sqrt{\frac{\pi}{4k}} \frac{\exp{(i\pi(\gamma+1)/2)}}{\cos(\pi\gamma)}
\; ,
\label{Aconstant}
\ENA
where we have used the asymptotic expansion, that for $x \to \infty$,
\[
J_\nu(x) \to \sqrt{\frac{2}{\pi x}} \cos \left[ 
x - (\nu+\frac12)\frac{\pi}{2} \right] \; .
\]
In the opposite limit of modes well outside the Hubble radius,
or at late times, with $(-k\eta) \to 0$, we get from \Eq{Asoln},
\EQ
{\cal A} \to k^{-1/2} \left[ c_1(\gamma) (-k\eta)^{\gamma}
+ c_2(\gamma) (-k\eta)^{1-\gamma} \right],
\label{finalA}
\EN
where
\EQA
c_1 &=& {\frac{\sqrt{\pi}}{2^{\gamma+1/2}}} 
\frac{e^{-i\pi\gamma/2}}{\Gamma(\gamma +\frac12)\cos(\pi\gamma)}, 
\nonumber \\
c_2 &=& {\frac{\sqrt{\pi}}{2^{3/2-\gamma}}} 
\frac{e^{i\pi(\gamma+1)/2}}{\Gamma(\frac32-\gamma)\cos(\pi\gamma)}, 
\label{confin}
\ENA
Here we have used the property that 
\[
J_\nu(x) \to \frac{x^\nu}{2^\nu \Gamma(\nu +1)}, \quad
{\rm as} \quad  x\to 0 \; .
\]
From \Eq{finalA} one sees that 
the $c_1$ term dominates for $\gamma \le 1/2$,
while $c_2$ term dominates for $\gamma \ge 1/2$.

As an aside, we note that the late time (small $k$) solution can also be got
for a more general $f$ from directly integrating \Eq{calAeq}
in the limit $k \to 0$. We get
\EQ
{\cal A} \to \bar{c}_1 f + \bar{c}_2 f \int \frac{d\eta}{f^2}.
\label{dirA}
\EN
If we substitute $f \propto a^\alpha$ and $a \propto \vert\eta\vert^{1+\beta}$
into \Eq{dirA}, we will recover the long wavelength solution given
in \Eq{finalA}.

\subsection{The generated magnetic and electric fields}

We can now calculate the spectrum of $\rho_B$ and $\rho_E$ in the
late time, super Hubble limit. 
Substituting \Eq{finalA} into \Eq{rhoB}, we get for the magnetic spectrum,
\EQA
\frac{d\rho_B}{d\ln{k}} &=& 
\frac{{\cal F}(n)}{2\pi^2} \ H^4
\ \left(\frac{k}{aH} \right)^{4+2n} \
\nonumber \\
&\approx& 
\ \frac{{\cal F}(n)}{2\pi^2} \ H^4
\ \left(-k\eta\right)^{4+2n},
\label{rhoBn}
\ENA
where $n=\gamma$ if $\gamma \le 1/2$ and $n =1-\gamma$ for
$\gamma \ge 1/2$, and
\EQ
{\cal F}(n) = 
\frac{\pi}{2^{2n+1}\Gamma^2(n +\frac12)\cos^2(\pi n)}\; . 
\label{calFn}
\EN
During slow roll inflation, the Hubble parameter $H$
is expected to vary very slowly, and thus most of the
evolution of the magnetic spectrum is due to the 
$(-k\eta)^{4+2n}$ factor.
One can see that the property of 
scale invariance of the spectrum (with
$4 + 2n = 0$), and having $\rho_B \sim a^0$ go together,
and they require either $\gamma = 3$ or $\gamma = -2$.

We can also calculate the electric field spectrum in a very
similar manner. We first find $({\cal A}/f)'$ from
\Eq{Asoln}, using the identities
\[
J_\nu' - \frac{\nu}{x} J_\nu = - J_{\nu + 1}, \quad
J_\nu' + \frac{\nu}{x} J_\nu = - J_{\nu - 1}.
\]
and then take the limit $(-k\eta) \to 0$. Substituting the result into
\Eq{rhoE} gives
\EQA
\frac{d\rho_E}{d\ln{k}} &=& 
\frac{{\cal G}(m)}{2\pi^2} \ H^4
\ \left(\frac{k}{aH} \right)^{4+2m} \
\nonumber \\
&\approx& 
\ \frac{{\cal G}(m)}{2\pi^2} \ H^4
\ \left(-k\eta\right)^{4+2m},
\label{rhoEn}
\ENA
where now $m=\gamma+1$ if $\gamma \le -1/2$ and $m =-\gamma$ for
$\gamma \ge -1/2$, and
\EQ
{\cal G}(m) = 
\frac{\pi}{2^{2m+3}\Gamma^2(m +\frac32)\cos^2(\pi m)}\; . 
\label{calGn}
\EN
Thus having a scale invariant magnetic spectrum implies
that the electric spectrum is not scale invariant, and
in addition can vary strongly with time. For example
if $\gamma =3$, then $(4+2m) = -2$, although $(4+2n) =0$.
In this case, at late times as $(-k\eta) \to 0$, the electric field
increases rapidly, with $\rho_E \propto (-k\eta)^{-2} \to \infty$.
There is then the danger of its energy
density exceeding the energy density in the universe during inflation
itself, unless the scale of inflation (or the value of $H^4$ is
sufficiently small. Such values of $\gamma$ are strongly
constrained by the back reaction on the background expansion
they imply (Martin and Yokoyama, 2008).

On the other hand consider the case near $\gamma=-2$. In this case
the magnetic spectrum is scale invariant, and at the same time
$(4+2m) = 2$, and so the electric energy density goes as
$(-k\eta)^2 \to 0$ as $(-k\eta) \to 0$. Thus these values
of $\gamma$ are acceptable for magnetic field generation
without severe back reaction effects.

We now discuss the evolution of the field after inflation.
Post inflationary reheating is expected to
convert the energy in the inflaton field to radiation
(which will include various species of relativistic charged particles).
For simplicity let us assume this reheating to be instantaneous.
After the universe becomes radiation dominated its conductivity
($\sigma$) becomes important. Indeed Turner and Widrow (1988)
showed that the ratio $\sigma/H \gg 1$. In order to take into account this
conductivity, one has to reinstate the interaction term in the
EM action, given in \Eq{emaction}. Further, as the inflaton has
decayed, we can take $f$ to have become constant with time
and settled to some value $f_0$.
Varying the action with respect to $A_\mu$ now gives
\[
F^{\mu\nu}_{; \ \nu} = \frac{4\pi J^\mu}{f_0^2}
\]
The value of $f_0$ thus
goes to renormalize the value of electric charge 
$e$ to be $e_N = e/f_0^2$.
This aspect raises an additional potential
problem 
for values of $\gamma \approx -2$, which has been
recently emphasized by Demozzi, Mukhanov and Rubinstein (2009)
(DMR).

Suppose the inflationary expansion is almost exponential with
$\beta =-2$, then for $\gamma \approx -2$, we have 
$\alpha = \gamma/(1+\beta) \approx 2$.
This implies that the function $f =f_i (a/a_i)^2$ increases
greatly during inflation, from its initial value of $f_i$ at $a=a_i$.
Thus if we want $f_0 \sim 1$ at the end of inflation, 
then at early times $f_i \ll f_0$ and the renormalized charge
at these early times $e_N  = e/f_i^2 \gg e$. DMR argue 
that one is then in a strongly coupled
regime at the beginning of inflation where such a theory
is not trustable. There is however the following naive caveat to the
above argument: Suppose one started with a weakly coupled theory where
$f_i \sim 1$. Then at the end of inflation $f_0 \gg f_i$, and so
the renormalized charge $e_N \ll e$. Such a situation
does not seem to have the problem of strong coupling raised by DMR;
however it does leave the gauge field extremely weakly coupled
to the charges at the end of inflation.
This also means that even if $\rho_B$ is large, the magnetic
field strength itself as deduced from \Eq{magenergy} is 
$B^iB_i = 8\pi \rho_B/f_0^2 \ll 8\pi\rho_B$. Ideas to sort out this
difficulty need to be explored, whether for example  
one can relax a large $f_0$ back to $f_i$, 
without now re-generating strong electric fields.

Let us proceed by assuming that we have absorbed $f_0^2$
into $e$. In the conducting plasma which obtains after reheating,
the current density will be given by the Ohm's law of \Eq{relohm}.
The fluid velocity at this stage is expected to be that of
the fundamental observers, i.e. $w^\mu = u^\mu$.
Thus the spatial components $J^i = \sigma E^i = -g^{ij}\dot{A_j}$.
Let us assume that the net charge density is negligible and
thus neglect gradients in the scalar potential $A_0$.
Then the evolution of the spatial components of the 
vector potential is given by
\EQ
\ddot{A}_i +(H + 4\pi\sigma) \dot{A}_i
 -\partial_j\partial^j A_i = 0 \; .
\label{Aevolsig}
\EN
We see that any time dependence in $A_i$ is damped out on
the inverse conductivity time-scale. To see this explicitly,
consider modes which have been amplified during inflation
and hence have super Hubble scales $k/(aH) \ll 1$.
Also let us look at the high conductivity limit
of $\sigma/H \gg 1$. Then \Eq{Aevolsig}
reduces to 
\[
\ddot{A}_i + 4\pi\sigma \dot{A}_i = 0 
\]
whose solution is given by
\EQ
A_i = \frac{D_1(\xx)}{4\pi \sigma} e^{-4\pi\sigma t}
+ D_2(\xx)\; .
\EN
We see that the $D_1$ term decays exponentially on a time-scale of
$(4\pi\sigma)^{-1} \ll (1/H)$. This leaves behind a constant
(in time) $A_i = D_2(\xx)$. Thus the electric field $E_i =0$, 
and the high conductivity of the plasma has led to the shorting out
of the electric field. Note that the time scale 
in which the electric field decays does not depend on the scale 
of the perturbation, that is the $\sigma$ dependent damping term
in \Eq{Aevolsig} has no dependence on spatial derivatives.
As far as the magnetic field is concerned, \Eq{EBinA} shows that
$B_i \sim 1/a$ when $A_i = D_2(\xx)$. Therefore $\bar{B}_i \sim 1/a^2$,
as expected when the magnetic field is frozen into the highly conducting plasma.

Let us now make a numerical estimate of the 
strength of the magnetic fields generated in the scale invariant case.
For both $\gamma =-2$ and $\gamma = 3$, we have
from \Eq{rhoBn} and \Eq{calFn},  
\EQ
\frac{d\rho_B}{d\ln{k}} 
\approx \frac{9}{4\pi^2} H^4 
\;.
\label{bfin}
\EN
Cosmic Microwave Background 
limits on the amplitude of scalar perturbations
generated during inflation, give an upper limit on $H/M_{pl} \sim 10^{-5}$
(cf. Bassett, Tsugikawa and Wands, 2006). Here
$M_{pl} = 1/\sqrt{G}$ is the Planck mass. The magnetic energy
density decreases with expansion 
as $1/a^4$, and so its present day value
$\rho_B(0) = \rho_B (a_f/a_0)^4$, where $a_f$ is the scale factor
at end of inflation, while $a_0$ is its present day value.
Let us assume that the universe transited to radiation domination
immediately after inflation and use entropy conservation,
that is the constancy of $g T^3 a^3$ during its evolution, where
$g$ is the effective relativistic degrees of freedom and $T$ the temperature
of the relativistic fluid.
We get
\[
\frac{a_0}{a_f} \sim \frac{g_f^{1/12}}{g_0^{1/3}} \frac{H^{1/2} M_{pl}^{1/2}}{T_0}
\left(\frac{90}{8\pi^3}\right)^{1/4}
\] 
Taking $g_f \sim 100$, gives $(a_0/a_f) \sim 10^{29} (H/10^{-5} M_{pl})^{1/2}$, 
This leads to an estimate the present day value of
the magnetic field strength, $B_0$ at any scale,
\EQ
B_0 \sim 5 \times 10^{-10} {\rm G} 
\left(\frac{H}{10^{-5} M_{pl}}\right)
\; .
\label{Bstrength}
\EN
Thus interesting field strengths can in principle be created
if the parameters of the coupling function $f$ are
set appropriately and the problems highlighted by
DMR can be circumvented.
Note that the strength of the generated field
is sensitive to even slight departures from scale invariance.
Suppose $\gamma = -2+\epsilon$, with $\epsilon \ll 1$, then 
\EQ
\frac{d\rho_B}{d\ln{k}}
\approx \frac{9}{4\pi^2} H^4
\left(\frac{k}{aH}\right)^{2\epsilon}.
\label{othspec}
\EN
valid for $(k/aH) \ll 1$. Assuming a radiation dominated universe
immediately after inflation, and matter domination 
from a redshift $z_{eq} \sim 3300$, we estimate
\[
(k/aH) \sim 3 \times 10^{-24} 
\left(\frac{k}{1 \ h \ {\rm Mpc}^{-1}}\right)
\left(\frac{H}{10^{-5} M_{pl}}\right)^{1/2}
\]
Thus at galactic scales of $k = 1 h$ Mpc$^{-1}$, 
$B_0$ will be smaller or larger by a factor of $\sim 10^{-5}$,
if one takes $\epsilon=\pm 0.2$ respectively.
This shows the sensitivity of the magnetic field amplitude to small
changes in the parameters of any generation model, as mentioned earlier.

\section{Discussion}

We end with a few comments. We have emphasized the use
of tetrads in defining properly behaved magnetic and electric fields.
Such a procedure is also followed when using with the $3+1$
formalism in the context of black-hole electrodynamics
(cf. Macdonald et al., 1986). Thus it seems to us somewhat surprising
why this does not usually get mentioned when dealing with
MHD in the context of cosmology.
Regarding magnetic field generation during inflation,
we have a paradigm, but no compelling model as yet.
Clearly more work is required to find such a model,
which at the same time avoids the problems of back reaction,
the strong coupling problem mentioned by DMR, 
or the strong decay of coupling constants.

Another possibility is generation of 
primordial fields is during a later phase transition,
like the Electro-Weak or the quark-hadron transition. Here the main
problem is that the generated field typically has a tiny correlation
scale, less than the Hubble radius $H^{-1}$, at the epoch
of the phase transition, unless magnetic helicity is
also generated. Interestingly there are several ideas for such
helicity generation during the Electro-weak phase transition 
(cf. Vachaspati, 2001; Diaz-Gil et al., 2008; Copi et al., 2008). 
Magnetic energy decay conserving helcity can then lead
to an inverse cascade and larger coherence scales, 
as first emphasized in the
early universe context by Brandenburg, Enquist and Olesen (1996)
(see also Christensson, Hindmarsh and Brandenburg, 2001;
Banerjee and Jedamjik, 2005), but that is a story for another review.

\acknowledgements
It is a pleasure to contribute to the special issue celebrating 
Axel Brandenburg's 50th birthday. I have had many
enjoyable collaborations with Axel, although not yet on the
topic of this review, and wish him all the best for the 
next 50 years worth of fruitful research. I also thank John Barrow, T. Padmanabhan,
Anvar Shukurov and L. Sriramkumar for several useful discussions, 
and an anonymous referee for useful comments.
\newpage

\end{document}